\documentclass[review]{elsarticle}

\usepackage{hyperref}
\usepackage{amsmath,graphicx, amssymb,tabularx}
\usepackage{subcaption}
\usepackage[font=small,skip=2pt]{caption}
\usepackage{boxedminipage}
\usepackage{algorithm}
\usepackage{algpseudocode}
\usepackage{algorithm}
\usepackage{xcolor}
\usepackage{mathtools}
\usepackage{comment}

\def\x{{\mathbf x}}

\journal{}

\usepackage{caption}
\begin{document}

\begin{frontmatter}

\title{Low radiation tomographic reconstruction with and without template information}

\author[mymainaddress,mysecondaryaddress,mythirdaddress]{Preeti Gopal}

\author[mymainaddress]{Sharat Chandran}

\author[mysecondaryaddress]{Imants Svalbe}
\author[mymainaddress]{Ajit Rajwade}

\address[mymainaddress]{Department of Computer Science and Engineering, IIT Bombay}
\address[mysecondaryaddress]{School of Physics and Astronomy, Monash University}
\address[mythirdaddress]{IITB-Monash Research Academy}
\address{\{preetig,sharat,ajitvr\}@cse.iitb.ac.in,imants.svalbe@monash.edu}

\begin{abstract}

  Low-dose tomography is highly preferred in medical procedures for its reduced radiation risk when compared to standard-dose Computed Tomography (CT). However, the lower the intensity of X-rays, the higher the acquisition noise and hence the reconstructions suffer from artefacts. A large body of work has focussed on improving the algorithms to minimize these artefacts. In this work, we propose two new techniques, rescaled non-linear least squares and  Poisson-Gaussian convolution, that reconstruct the underlying image making use of an accurate or near-accurate statistical model of the noise in the projections. We also propose a reconstruction method when prior knowledge of the underlying object is available in the form of templates. This is applicable to longitudinal studies wherein the same object is scanned multiple times to observe the changes that evolve in it over time. Our results on 3D data show that prior information can be used to compensate for the low-dose artefacts, and we demonstrate that it is possible to simultaneously prevent the prior from adversely biasing the reconstructions of new changes in the test object, via a method called ``re-irradiation''. Additionally, we also present two techniques for automated tuning of the regularization parameters for tomographic inversion.

\end{abstract}


\end{frontmatter}

\section{Introduction}
\label{sec:intro}
Reduction in radiation exposure is a critical goal, especially in CT of medical subjects~\cite{Brenner2007} and biological specimens~\cite{Howells2009}. One of the ways to reduce this radiation is to acquire projections from fewer views. An alternate way, which is the focus of this work, is to lower the strength (`dose') of X-ray beam. The CT imaging model that incorporates the strength of X-rays, $I_0$, is non-linear and non-deterministic and is given by:
\begin{equation}
\boldsymbol{y} \sim \rm{Poisson}(I_0\exp\{-\boldsymbol{\Phi x}\}) + \boldsymbol{\eta}
\label{eq:GaussyActualModel}
\end{equation}
 where $\boldsymbol{\eta}$ represents the zero mean additive Gaussian noise vector with a fixed signal-independent standard deviation $\sigma$, where $\boldsymbol{\Phi}$ is the sensing matrix which represents the forward model for the tomographic projections, and $\boldsymbol{x}$ is the underlying image representing the density values. The noise model for $\boldsymbol{y}$ is primarily Poisson in nature as this is a photon counting process~\cite{image_science}, and the added Gaussian noise is due to the thermal effects~\cite{Xu2009}. This Poisson-Gaussian noise model is quite common in optical or X-ray based imaging systems, but we consider it here explicitly for tomography, where it induces a non-linear inversion problem.  Specifically, the $i^{th}$ index (for bin number and projection angle) in the measurement vector $\boldsymbol{y}$ is given as:
$y_i \sim \textrm{Poisson}(I_0 \exp\{-\boldsymbol{\Phi^i x}\}) + \eta_i $, where $\boldsymbol{\Phi^i}$ is the $i^{th}$ row of the sensing matrix $\boldsymbol{\Phi}$. The major effect of low-dose acquisition is the large magnitude (relative to the signal) of Poisson noise due to the low strength of X-ray beam. This is because the Signal-to-Noise-Ratio (SNR) of Poisson noise with mean $\lambda$ and variance $\lambda$ is given by $\frac{\lambda}{\sqrt\lambda}=\sqrt\lambda$. Due to the inherently low SNR, traditional low dose reconstructions are noisy. 

\section{Previous Work}
\label{sec:previous_work}
Modelling of Poisson noise and recovery of images also finds applications in areas outside of CT.~\cite{ZHANG2017} recovered images from Poisson-noise corrupted and blurred images using alternating direction method of multipliers(ADMM). 
Low-dose imaging and reconstruction (with dense projection view sampling) has been more widely studied than the few-views imaging. This is probably because the former does not involve a strategy for selection of the set of view angles, which in itself is an active field of research~\cite{barkan17,fischer16,andrei14}. For long, almost all of the commercial CT machines used FBP~\footnote{Filtered Backprojection} as the standard reconstruction technique~\cite{Pan2009}. Only recently are the iterative techniques being deployed for commercial use~\cite{PhilipsNews2013}. The power of iterative routines was reinforced by~\cite{Fujita2017}, where it was proved that iterative reconstructions from \textit{ultra-low} dose~\footnote{Typically, low-dose imaging is performed at 120 kVp and 30 mAs beam current, and ultra-low dose imaging is performed at 80-100 kVp and 20-30 mAs beam current settings.} CT are of similar quality to those of FBP reconstructions from \textit{low}-dose CT. Here, a commercial forward projected model-based algorithm was deployed and compared with FBP.

 Among the other iterative methods,~\cite{Kim2017} presented a technique that minimizes log-likelihood of the Poisson distribution and a patch-based spatially encoded non-local penalty.~\cite{Lyu2019} used a smoothness prior along with data-fidelity constraint and solved using ADMM. In order to improve the reconstruction further, various prior-based and learning-based methods have also been explored in literature. In these techniques, properties of available standard-dose CT images influence low-dose reconstruction of the test (i.e., the object which needs to be reconstructed from the current set of new tomographic projections). One such technique was described by~\cite{Zheng2016},  wherein the iterative reconstruction was formulated as a penalized weighted least squares problem with a pre-learned sparsifying transform. While the weights were set manually, the sparsifying transform was learned from a database of regular-dose CT images. Another technique presented by~\cite{Jia2016} clustered overlapping patches of previously scanned standard-dose CT images using Gaussian Mixture Model (GMM). The texture of the prior was learned for each cluster. Following this, patches from a pilot reconstruction of the test were classified using the learned GMM and depending on the class, the corresponding texture priors were imposed on patches of the reconstructed test image. The limitation here is-- patches that correspond to new changes between the test and the templates will also be influenced by some inappropriate texture of patches from prior. \cite{Zhang2016} solved a cost function with L1 norm for imposing similarity to a learned dictionary. They concluded that the number of measurements needed is progressively less for each of the four methods: Simultaneous Algebraic Reconstruction Technique (SART)~\cite{sart}, Adaptive Dictionary based Statistical Iterative Reconstruction (ADSIR)~\cite{Xu2012}, Gradient Projection Barzilai Borwein (GBPP)~\cite{Park2012} and their method (L1-DL), in the same order.~\cite{Jiao2013} used edge-based priors to reconstruct normal-dose CT along with Compressed Sensing (CS) sparsity prior. An iterative method~\cite{Jing2012} in a related area (electrical impedance tomography) reconstructs using Split Bregman algorithm for L1 minimization. \textit{None of these methods explore optimizing a log-likelihood based cost-function that accurately reflects the Poisson-Gaussian noise statistics. In addition, they do not address the issue of the prior playing a role in the reconstruction of parts of the test that are dissimilar to the parts of the prior, which is undesirable.} In contrast, this work focuses on  applying a computationally fast  global prior on only those regions of the test that are similar to the prior. 

 Lately, artificial neural networks have also been designed for low-dose reconstruction.~\cite{Wu2017} proposed one such neural network to learn features of the image that is later imposed along with data-fidelity during iterative reconstruction.~\cite{Hong2019} showed that deep neural network based reconstructions are faster than iterative reconstructions for comparable reconstruction quality. \textit{All of these neural-network based techniques need large amount of data. This can be challenging in longitudinal studies where usually only a few of the previous scans of the same object are available.} Hence, this paper focuses on analytical iterative techniques.

 We also present a technique for parameter selection. Most techniques in literature tune the parameters omnisciently. A recent work~\cite{chang2019}  used the L-curve method in which data-fidelity residue is plotted against regularization norm. The parameter can then be selected based on the performance required for the application at hand. However, this method does not utilize the available information about noise statistics in low-dose imaging. In this work, we use the noise-model for the purpose of automated parameter selection.   

 \section{Contributions}
 
 This paper discusses the following:
\begin{enumerate}
\item How the quality of reconstruction is affected by the manner in which Poisson and Gaussian noise are modelled within iterative routines.  
\item How a prior of the object being scanned can be effectively used to compensate for the noisy low dose measurements, while simultaneously identifying genuine structural changes between the currently scanned objects and its priors, and preventing undesirable influence of the prior in those regions. We also propose a technique called re-irradiation which improves the reconstruction quality in these regions of change at the cost of a small amount of added radiation.
  \item  In addition, most of the iterative schemes involve a cost function with a data-fidelity term, a sparsity term and a data-prior term.  We discuss a systematic way to tune the parameters involved with these terms.
\end{enumerate}
Specifically, this work presents a few new reconstruction methods and their comparison with existing methods, each of which model noise in a slightly different way. In addition to this, a technique for detecting new changes (i.e., differences between the test and templates) \textit{directly} in the measurement space is presented. This is applied for prior-based reconstruction in longitudinal studies.

Sec.~\ref{sec:reconstruction_without_prior} describes two new techniques and its comparison with a few methods in literature. Sec.~\ref{sec:low_dose_reconstruction_with_prior} describes the new prior-based low-dose reconstruction and its validation on 3D tomographic data. 
In Sec.~\ref{sec:tuning_hyper_parameters}, we illustrate a systematic technique to parameter-tuning. Finally, key results are summarized in Sec.~\ref{sec:conclusions}

 \section{Reconstruction without prior}
 \label{sec:reconstruction_without_prior}
A good low-dose reconstruction technique should make optimal use of noise statistics as well as appropriate signal priors. Most techniques will involve minimizing a cost function of the following form: $J(\boldsymbol{x};\boldsymbol{y,\Phi}) = DF(\boldsymbol{y}|\boldsymbol{\Phi x}) + \lambda R(\boldsymbol{x})$. Here the first term involves a data-fidelity cost, and may possibly (though not necessarily) be expressed by the negative log-likelihood of $\boldsymbol{y}$ given $\boldsymbol{\Phi}$ and $\boldsymbol{x}$ (i.e., by $-\log p(\boldsymbol{y}|\boldsymbol{\Phi,x}))$. Other alternatives could include a simple least squares term $\|\boldsymbol{y} - \boldsymbol{\Phi x}\|^2_2$, or a weighted version of the same. The second term $R(\boldsymbol{x})$ is a regularizer (weighted by the regularization parameter $\lambda$) representing prior knowledge about $\boldsymbol{x}$. This could be in the form of the well-known total variation prior $TV(\boldsymbol{x}) = \sum_{i,j}\sqrt{(x(i+1,j) - x(i,j))^2 + (x(i,j+1) - x(i,j))^2}$ or penalty on the $\ell_1$ norm of the coefficients $\boldsymbol{\theta}$ in a sparsifying basis $\boldsymbol{\Psi}$ where $\boldsymbol{x} = \boldsymbol{\Psi\theta}.$ Such cost functions are minimized by iterative shrinkage and thresholding algorithms such as ISTA. However, ISTA by itself is known to have slow convergence (as discussed in Sec.3 of~\citep{FISTA}). Hence, faster methods such as the Fast Iterative Soft Thresholding Algorithm (FISTA)~\citep{FISTA} may be used, which is the method adopted in this paper. Below are some of the existing reconstruction methods, or intuitive variants thereof, and two new proposed techniques.

\subsection{Post-log Compressed Sensing (CS)}\label{subsubsec:post_log_CS}  
\begin{figure}
\centering
  \includegraphics[width=0.4\linewidth]{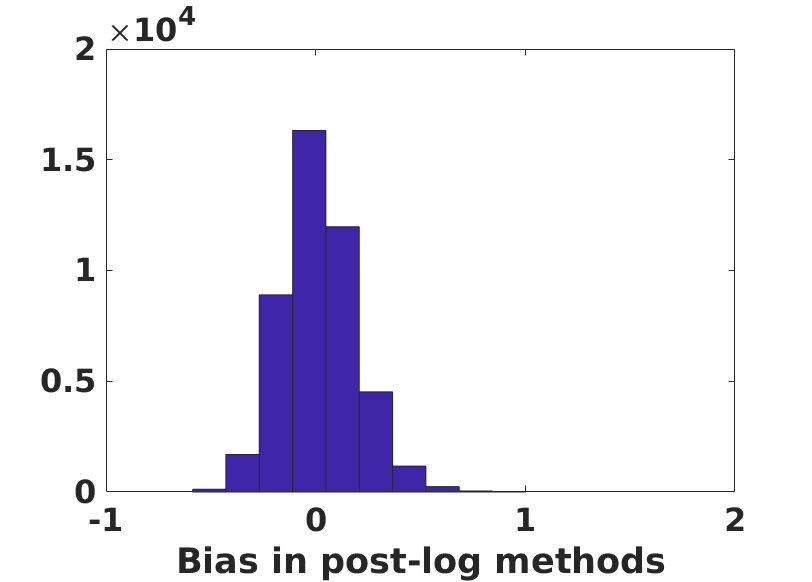}
  \caption[Statistical bias in post-log methods]{Histogram of statistical bias in post-log methods. The bias is computed as $(\boldsymbol{y}_0  - \boldsymbol{\Phi\Psi\theta})$, where $\boldsymbol{y}_0$ refers to linearized post-log measurements. Here, the added Gaussian noise had a mean value of $0$ and $\sigma = 0.01 \times$ average Poisson-corrupted projection value. The fact that every bin has a different bias, but  is shifted by a constant $\boldsymbol{\epsilon}$ is problematic. This results in poor reconstructions, as shown in a later Sec.~\ref{sec:comparison}.}
  \label{fig:histogram}
\end{figure}

A preliminary approach is to ignore the presence of Poisson noise and apply traditional CS reconstruction after linearizing the measurements~\citep{Hou2014}. The latter process is performed by computing the logarithm of the acquired measurements. The linearized measurements $\boldsymbol{y_0}$ are given by
$\boldsymbol{y_0} = - \log\Big(\frac{\boldsymbol{y}+\boldsymbol{\epsilon}}{I_0}\Big) \boldsymbol{ \Phi\Psi\theta}$,
where $\boldsymbol{\epsilon}$ is a small positive constant added to the measurements to make them all positive and thus suitable for linearizing by applying a logarithm. For practical purposes, if $\rm{min}(\boldsymbol{y})$ is zero or negative, $\boldsymbol{\epsilon}$ is set to $-\rm{min}(\boldsymbol{y}) + 0.001$. The cost function is given by
\begin{equation}
J_{PL-CS}(\boldsymbol{\theta}) = \lVert\boldsymbol{y_0} - \boldsymbol{\Phi\Psi\theta}\rVert_2^2 + \lambda\lVert\boldsymbol{\theta}\rVert_1 \textrm{, subject to } \boldsymbol{\Psi \theta} \succeq \boldsymbol{0}
\label{eq:Post-log2}
\end{equation}
 $J_{PL-CS}$ is minimized by $l1-ls$ solver~\citep{l1ls}. This method is however not true to the Poisson-Gaussian statistics and suffers from an inherent statistical bias (as seen in Fig.~\ref{fig:histogram}) as it is a so-called `post-log' method. The bias arises because for any non-negative random variable $X$, we have $\log(E[X])\geq E(\log(X))$ as per Jensen's inequality. Another way of viewing this is that the noise in $\boldsymbol{y}_0$ (i.e. post-log) is being treated as if it were Gaussian with a constant variance). This is not true except at very high intensity ($I_0$) value. The adverse effects of computing post-log measurements is also discussed in~\citep{Lin2016}.

\subsection{Non-linear Least Squares with CS} 
\label{subsubsec:NLLSCS}

An intuitive way to modify the previous cost  $J_{PL-CS}$ is by allowing the data fidelity cost to mimic the non-linearity inherent in the acquisition process. The cost function is then given by
\begin{equation}
J_{NL-CS} =  \lVert\boldsymbol{y} - I_0e^{-\boldsymbol{\Phi\Psi\theta}}\rVert_2^2 + \lambda\lVert\boldsymbol{\theta}\rVert_1  \text{, subject to } \boldsymbol{\Psi\theta} \succeq\boldsymbol{0}
\label{eq:NL-CS}
\end{equation}
The FISTA routine is used for this minimization. Since the attenuation constant of an object is never negative, a non-negativity constraint is imposed on $\boldsymbol{\Psi\theta}$. It can be seen that this cost function is non-convex in $\boldsymbol{\theta}$. Moreover, it treats all measurements as though they had the same noise variance, which is not true of Poisson settings.

\subsection{Filtered Backprojection}
In this technique, the classic filtered backprojection is applied on the linearized measurements: 
$\boldsymbol{y_0} = - \log\frac{\boldsymbol{y}+\epsilon}{I_0}
= \boldsymbol{\Phi x}$.
The slice or volume $\boldsymbol{x}$ is then reconstructed from the linearized measurements by filtered backprojection (FBP) in case of parallel beam projections or FeldKamp David Kress (FDK) algorithm~\citep{FDK} in case of cone beam projections. This method is called the post-log FBP. While it is computationally efficient, it suffers from a statistical bias for the same reasons as post-log CS, as described in~\ref{subsubsec:post_log_CS}. The performance of post-log FBP has been extensively compared with iterative schemes in~\citep{Pontana2011},\citep{Haiyan2013},\citep{Koyama2014} and the latter has been found to be well suited for low-dose reconstructions~\citep{Willemink2019}.

\subsection{Negative Log Likelihood-Poisson with CS}
This technique accounts for only the Poisson noise (ignoring the Gaussian part) and searches for a solution that minimizes the negative log-likelihood of the observed measurements. Given $m$ measurements, the likelihood of $\boldsymbol{\theta}$ is defined as
\begin{equation}
L(\boldsymbol{\theta|y})\coloneqq P_Y(\boldsymbol{Y=y|\theta})= \prod_{i=1}^{m} \frac{e^{-a_i}a_i^{y_i}}{y_i!}
\end{equation}
where $a_i = I_0e^{-\boldsymbol{(\Phi\Psi\theta)}_i}$. Thus, the negative log likelihood of $\boldsymbol{\theta}$ is given by
\begin{equation}
\begin{split}
 -\log(P(\boldsymbol{y|\theta})) &= \sum_i(a_i - y_i\log a_i + \log(y_i!))\\
&=\sum_i(I_oe^{-(\boldsymbol{\Phi\Psi\theta})_i} - y_i(\log(I_0) - (\boldsymbol{\Phi\Psi\theta})_i) + \log(y_i!))
\end{split}
\label{eq:NegLogLP1}
\end{equation}
The cost function combines the likelihood and the CS term as shown below.
\begin{equation}
 J_{NLL-P}(\boldsymbol{\theta})=\sum_i(I_oe^{-(\boldsymbol{\Phi\Psi\theta})_i} - y_i(\log(I_0) - (\boldsymbol{\Phi\Psi\theta})_i) + \lambda\lVert\boldsymbol{\theta}\rVert_1  \textrm{, subject to } \boldsymbol{\Psi \theta} \succeq \boldsymbol{0}
\label{eq:NegLogLP2}
\end{equation}


\subsection {Negative Log Likelihood-Poisson-Gaussian with CS}
\label{sec:negative log-likelihood Poisson-Gaussian}
A natural extension of the earlier method is one wherein both the Poisson and Gaussian noise processes are accounted for in the design of the cost function. Here, given the measurements, the solution that minimizes the sum of negative likelihood terms of both Poisson and Gaussian noise models, is selected. Let $\boldsymbol{V}$ denote the Poisson random variable, i.e. $\boldsymbol{y = v + \eta}$. As seen earlier, the Poisson likelihood of $\boldsymbol{\theta}$ is given by
\begin{equation}
L(\boldsymbol{\theta|v}) \coloneqq P_V(\boldsymbol{V=v|\theta})= \prod_{i=1}^{m} \frac{e^{-a_i}a_i^{v_i}}{v_i!} 
\end{equation}
where $a_i = I_0e^{-\boldsymbol{\Phi\Psi x}}$.  Poisson negative log-likelihood of $\theta$ is given by
\begin{equation}
\begin{split}
 -\log(P_V(\boldsymbol{V=v|\theta})) &= \sum_i(a_i - v_i\log a_i + \log(v_i!))\\
&=\sum_i(I_oe^{-(\boldsymbol{\Phi\Psi\theta})_i} - v_i(\log(I_0) - (\boldsymbol{\Phi\Psi\theta})_i) + \log(v_i!))
\end{split}
\label{eq:NegLogLP1}
\end{equation}
Next, if the assumed Gaussian noise has a variance of $\sigma^2$, then Gaussian likelihood of $\sigma$ is given by
\begin{equation}
L(\sigma|\boldsymbol{\eta}) \coloneqq P_E(\boldsymbol{E=\eta}|\sigma)
 = P((\boldsymbol{y-v})|\sigma)\\
= \prod_{i=1}^{m} e^{-\frac{(y_i-v_i)^2}{2\sigma^2}}
\end{equation}
The Gaussian negative log-likelihood of $\sigma$ is given by
\begin{equation}
 -\log(P(\boldsymbol{y-v})|\sigma)  = \sum_i\frac{(y_i-v_i)^2}{2\sigma^2}
\label{eq:NegLogLG1}
\end{equation}
 We minimize the sum of the two negative log-likelihoods:
\begin{equation}
\begin{split}
 J_{PG-NLL}(\boldsymbol{\theta,v})&=\sum_i(I_oe^{-(\boldsymbol{\Phi\Psi\theta})_i} - v_i(\log I_0 - (\boldsymbol{\Phi\Psi\theta})_i) + \log(v_i!)\\ &+ \frac{(y_i-v_i)^2}{2\sigma^2}) + \lambda\lVert\boldsymbol{\theta}\rVert_1  \textrm{, subject to } \boldsymbol{\Psi \theta} \succeq \boldsymbol{0}
\end{split}
\label{eq:NegLogLP2}
\end{equation}
 $\boldsymbol{\theta}$ and $\boldsymbol{v}$ are solved for alternately. Note that $\boldsymbol{v}$ is integer-valued,  but a typical gradient-based method will not restrict $\boldsymbol{v}$ to remain in the domain of integers. For computational convenience, $\boldsymbol{v}$ needs to be `softened' to real values. Consequently $\log (v_i!)$ must be replaced by the gamma function. 

This cost function is non-convex. However it can be shown to be bi-convex, i.e., it is convex in $\boldsymbol{\theta}$ if $\boldsymbol{v}$ is kept fixed and vice versa. Such a cost-function was used in~\citep{xie2017} as a method of pre-processing/denoising of projections prior to tomographic reconstruction. In contrast, we directly use it as a data-fidelity term for tomographic reconstruction. This appears more principled because denoising of a projection induces some `method noise' which cannot be accurately modelled and which may affect subsequent reconstruction quality.

\subsection{Proposed Rescaled non-linear Least Squares (RNLLS) with CS} 
\label{sec:rescaledNLLSCS}
This new method integrates Poisson noise model into the technique described in Sec.\ref{subsubsec:NLLSCS}. Since, the variance of a Poisson random variable is proportional to its mean, the variance of $\boldsymbol{y}$ is directly proportional to $I_0\exp(-\boldsymbol{\Phi\Psi\theta})$. Hence the data-fidelity cost must be rescaled as shown below: 
\begin{equation}
J_{RNLLS}(\boldsymbol{\theta}) = \sum_{i=1}^m\frac{(y_i - I_0e^{(-\boldsymbol{\Phi\Psi\theta})_i})^2}{I_0e^{(-\boldsymbol{\Phi\Psi\theta})_i}} + \lambda\lVert\boldsymbol{\theta}\rVert_1  \textrm{, subject to } \boldsymbol{\Psi \theta} \succeq \boldsymbol{0}
\label{eq:RescaledNLLSCS}
\end{equation}
Again, the cost is minimized using FISTA solver. This technique is in some sense similar to the Penalized Weighted Least Squares (PWLS) technique from~\citep{Fessler1994} which seeks to minimize
\begin{equation}
J_{PWLS}(\boldsymbol{\theta}) = \lVert\boldsymbol{W}(\boldsymbol{y - \Phi\Psi \theta})\rVert^2 + \lambda\lVert\boldsymbol{\theta}\rVert_1
\end{equation}
where $\boldsymbol{W}$ is a diagonal matrix of weights which are explicitly set (prior to running the optimization) based on the values in $\boldsymbol{y}$. In contrast to PWLS, in RNLLS, no weights are set as a prior. Rather the weights are equal to the underlying noiseless measurement values, and are explicitly inferred on the fly. In fact, a major motivation for our proposed technique is based on the fact that 
\begin{equation}
E\Bigg(\frac{[y_i - I_0\exp(-\boldsymbol{\Phi\Psi\theta})_i]^2}{I_0\exp(-\boldsymbol{\Phi\Psi\theta})_i}\Bigg) = Var\Bigg(\frac{[y_i - I_0\exp(-\boldsymbol{\Phi\Psi\theta})_i]}{I_0\exp(-\boldsymbol{\Phi\Psi\theta})_i}\Bigg) = 1
\end{equation}
This technique can be used for the case of Poisson-Gaussian noise as well, as in
\begin{equation}
J_{RNLLS-PG}(\boldsymbol{\theta}) = \sum_{i=1}^m\frac{(y_i - I_0e^{(-\boldsymbol{\Phi\Psi\theta})_i})^2}{I_0e^{(-\boldsymbol{\Phi\Psi\theta})_i} + \sigma^2} + \lambda\lVert\boldsymbol{\theta}\rVert_1  \textrm{, subject to } \boldsymbol{\Psi \theta} \succeq \boldsymbol{0}
\label{eq:RescaledNLLSCS_PG}
\end{equation}
We noticed that in~\citep{ding::sir}, tomographic reconstruction was performed by minimizing the following cost function:
\begin{equation}
J_{RNLLS-PG-log}(\boldsymbol{\theta}) = \sum_{i=1}^m\frac{(y_i - I_0e^{(-\boldsymbol{\Phi\Psi\theta})_i})^2}{I_0e^{(-\boldsymbol{\Phi\Psi\theta})_i} + \sigma^2} + \langle\log(I_0\exp(-\boldsymbol{\Phi\Psi\theta})_i + \sigma^2),1\rangle
\label{eq:RescaledNLLSCS_PG}
\end{equation}
which is inspired by the approximation of $\textrm{Poisson}(z)$ by $\mathcal{N}(z,z)$  and treating it as a maximum quasi-likelihood problem. On the other hand, the proposed method (RNLLS) can be interpreted as a weighted form of the well-known LASSO problem~\citep{lasso_book}. We also note that the cost function for RNLLS is convex in the case of Poisson noise, as shown in the supplemental material. In the case of Poisson-Gaussian noise, our numerical simulations reveal that the cost function is not convex in the worst case. However, this non-convexity did not affect the numerical results significantly. 


\subsection{Proposed Poisson-Gaussian Convolution}

This new technique models both the Poisson and Gaussian noise. It is based on the fact that if a random variable $Q$ is the sum of two random variables $R$ and $S$, then the density function of $Q$ is given by the convolution of the density functions of $R$ and $S$. This scheme has been used earlier~\citep{emilie2015} for image restoration from linear degradations such as blur, followed by Poisson-Gaussian corruption of the signal. In contrast, in CT, the measured signal is a non-linear function of the underlying image (i.e. its attenuation coefficients) as per Beer's law. Eq.~\ref{eq:beers_law_noisy} refers to the Beer's law along with the Poisson and Gaussian noise. The measurement is the sum of a Poisson random variable and a Gaussian random variable:
\begin{equation}
  \boldsymbol{y} \sim \textrm{Poisson}(\boldsymbol{a}) + \boldsymbol{\eta}
  \label{eq:beers_law_noisy}
\end{equation}
where $\boldsymbol{a}$ = $I_0 e^{-\boldsymbol{\Phi\Psi\theta}}$. The $i^{\textrm{th}}$ measurement is given as: $y_i \sim \textrm{Poisson}(a_i) + \eta_i$, where $a_i = I_oe^{-[\boldsymbol{\Phi \Psi \theta}]_i}$. The probability density of the $i^{th}$ measurement $y_i$ is given by the following convolution:
 \begin{equation}
 p_{y_i}(z_i) = \sum_{l=0}^{l=+\infty}\frac{e^{-a_i}a_i^l}{l!}\frac{1}{\sigma \sqrt{2\pi}} e^{-\frac{(z_i - l)^2}{2\sigma^2}} 
\end{equation}
The running variable does not take on negative values because the Poisson is a counting process and hence the corresponding random variable is always positive. Because all the $m$ measurements are independent (i.e., the noise in the sensor at any one pixel is independent of the noise at any other pixel on it), we have
\begin{equation}
p_{\boldsymbol{y}}(\boldsymbol{z}) = \prod_{i=1}^{i=m}\Big(\sum_{l=0}^{l=\infty}\frac{e^{-a_i}a_i^l}{l!}\frac{1}{\sigma \sqrt{2\pi}} e^{-\frac{(z_i - l)^2}{2\sigma^2}}\Big)
\end{equation}
The $\boldsymbol{\theta}$ that maximizes the above probability needs to be computed. This is equivalent to minimizing the negative log-likelihood of the above probability. Hence, our cost function $J_{conv}$ is given by
\begin{equation}
\begin{split}
J_{conv}(\boldsymbol{\theta}) &= - \log p_{\boldsymbol{y}}(\boldsymbol{z})\\
   &= \sum_{i=1}^{i=m}-\log \Big(\sum_{l=0}^{l=\infty}\frac{e^{-I_oe^{-[\boldsymbol{\Phi \Psi \theta}]_i}}(I_oe^{-[\boldsymbol{\Phi \Psi \theta}]_i})^l}{l!}\frac{1}{\sigma \sqrt{2\pi}} e^{-\frac{(z_i - l)^2}{2\sigma^2}}\Big)\\ &  + \lambda\lVert\boldsymbol{\theta}\rVert_1  \textrm{, subject to } \boldsymbol{\Psi \theta} \succeq \boldsymbol{0}
\end{split}
\end{equation}
Since $l!$ is computationally intractable for large $l$, it has been approximated using Stirling's approximation: $l! \sim \sqrt{2\pi l}\ \big(\frac{l}{e}\big)^l$.
Further, in order to make the optimization numerically feasible, the value that $l$ takes for a particular measurement $y_i$ is limited to the range $\textrm{max}(0,y_i- K \sigma)$ to $y_i + K \sigma$ where $K$ is an integer that is usually set to 3. It is assumed here that some estimate of the variance $\sigma^2$ of the Gaussian noise is already known. This is usually feasible by recording the values sensed by the detector during an empty scan (without any object), usually before the actual scan is taken. 
Among the methods discussed here, the ones that model both Poisson and Gaussian noise are non-convex. A few of the methods that model Poisson noise alone are convex and their convexity is proved in Sec.1 of~\cite{videos}.

\subsection{Results on comparison of different methods}
\label{sec:comparison}
\begin{figure}[!h]
\centering
    \begin{subfigure}[b]{0.25\linewidth}
        \includegraphics[width=\textwidth]{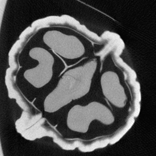}   
 \caption{walnut}
    \end{subfigure}
    \begin{subfigure}[b]{0.25\linewidth}
        \includegraphics[width=\textwidth]{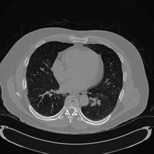}   
 \caption{colon}
    \end{subfigure}
       \caption[Test images]{Ground truth test slices used for comparison of low dose reconstruction techniques. A slice from (a)~\citep{walnut} dataset is of size $156\times156$, (b)~\citep{colon} dataset is of size $154\times154$}
\label{fig:low_dose_comparison_test_images}
\end{figure}

In order to compare the performance of various methods, 2D reconstructions of two datasets (Walnut and Colon CT) shown in Fig.~\ref{fig:low_dose_comparison_test_images} were computed for varying low-dose intensities. Reconstructions of two other datasets (Pelvis and Shoulder CT) are shown later in the supplemental material~\cite{videos}. Following are the details of the datasets and the conditions used for simulating low-dose imaging:
 The size of the image from Walnut dataset was $156\times156$, and the size of image from Colon CT dataset was $154\times154$.
 The sum of the intensity values for the Walnut and Colon dataset images were $75$ and $60$ respectively.
 Measurements were simulated using parallel beam geometry.
 The Cosine filter was applied for filtered backprojection. 
 While the number of projection views was large (200 views for all datasets) and kept constant, the beam strength $I_0$ was varied as follows: $I_0 = 20,40,80,160,320$ and $620$. 
 Based on the intensity (attenuation coefficients) of the images, the above values of $I_0$ correspond to a Poisson noise-to-signal ratio (i.e. average value of $1/\sqrt{\lambda}$) of $25\%$  for $I_0 = 20$, and $4.5\%$ for $I_0 = 620$, for both the datasets. 
 In addition, Gaussian noise of $0$ mean and variance equal to $2\%$ of average Poisson-corrupted measurement was added to measurements. The regularization parameter $\lambda$ was chosen omnisciently.


\begin{figure}[h!]
  \centering
    \begin{subfigure}[b]{0.49\linewidth}
        \includegraphics[width=\textwidth]{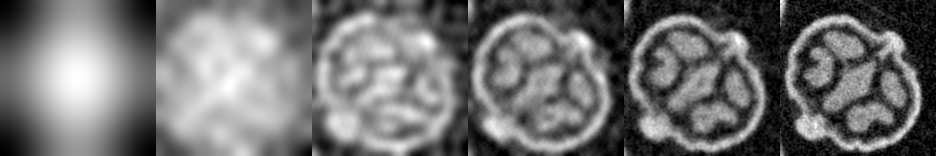}
\captionsetup{labelformat=empty}       
 \caption{Convolution}
    \end{subfigure}
    \begin{subfigure}[b]{0.49\linewidth}
        \includegraphics[width=\textwidth]{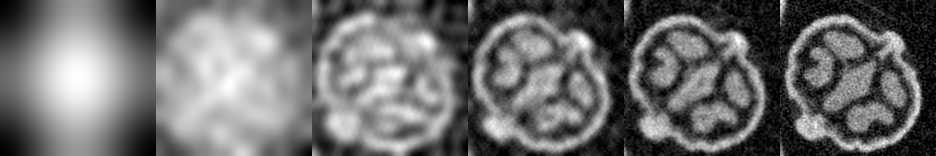}
\captionsetup{labelformat=empty}
        \caption{Log-Likelihood Poisson-Gaussian}
     \end{subfigure}
    \begin{subfigure}[b]{0.49\linewidth}
        \includegraphics[width=\textwidth]{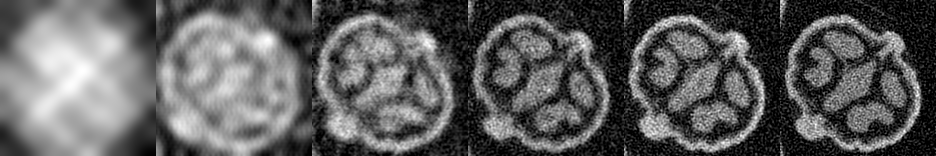}
\captionsetup{labelformat=empty}
        \caption{Rescaled Non-Linear Least Squares}
     \end{subfigure}
    \begin{subfigure}[b]{0.49\linewidth}
        \includegraphics[width=\textwidth]{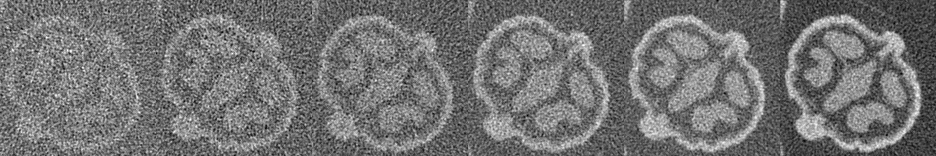}
\captionsetup{labelformat=empty}
        \caption{Post-Log FBP}
     \end{subfigure}
     \caption[Low-dose reconstruction on Walnut dataset]{2D Low-dose reconstructions of Walnut dataset for $I_0 = 20,40,80,160,320$ and $620$. Gaussian noise of $0$ mean and variance equal to $2\%$ of average Poisson-corrupted measurement was added to simulate the low-dose acquisition. The SSIM values are shown in Fig.~\ref{fig:low_dose_comparison_SSIM}.}
\label{fig:walnut_result}
\end{figure}


\begin{figure}[h!]
  \centering
    \begin{subfigure}[b]{0.49\linewidth}
        \includegraphics[width=\textwidth]{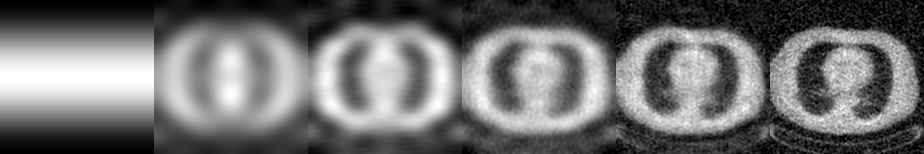}
\captionsetup{labelformat=empty}       
 \caption{Convolution}
    \end{subfigure}
    \begin{subfigure}[b]{0.49\linewidth}
        \includegraphics[width=\textwidth]{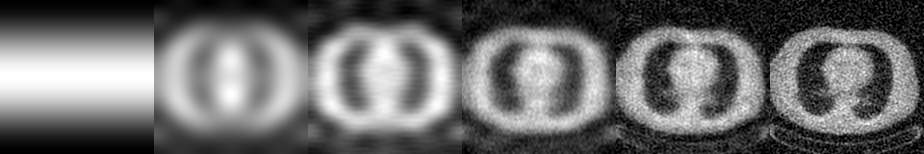}
\captionsetup{labelformat=empty}
        \caption{Log-Likelihood Poisson-Gaussian}
     \end{subfigure}
    \begin{subfigure}[b]{0.49\linewidth}
        \includegraphics[width=\textwidth]{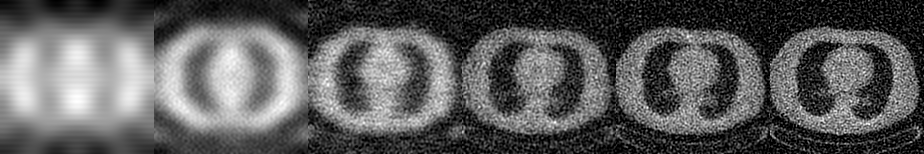}
\captionsetup{labelformat=empty}
        \caption{Rescaled Non-Linear Least Squares}
     \end{subfigure}
    \begin{subfigure}[b]{0.49\linewidth}
        \includegraphics[width=\textwidth]{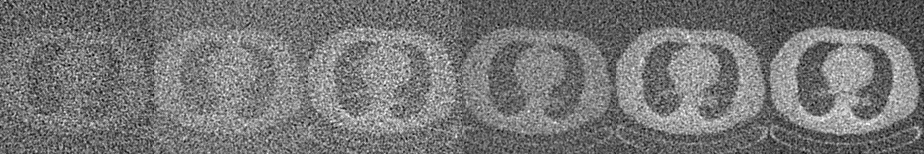}
\captionsetup{labelformat=empty}
        \caption{Post-Log FBP}
     \end{subfigure}
     \caption[Low-dose reconstruction on Colon dataset]{2D Low-dose reconstructions of Colon dataset for $I_0 = 20,40,80,160,320$ and $620$. Gaussian noise of $0$ mean and variance equal to $2\%$ of average Poisson-corrupted measurement was added to simulate the low-dose acquisition. The SSIM values are shown in Fig.~\ref{fig:low_dose_comparison_SSIM}.}
\label{fig:colon_result}
\end{figure}


\begin{figure}[h!]
  \centering
     \begin{subfigure}[b]{0.45\linewidth}
        \includegraphics[width=\textwidth]{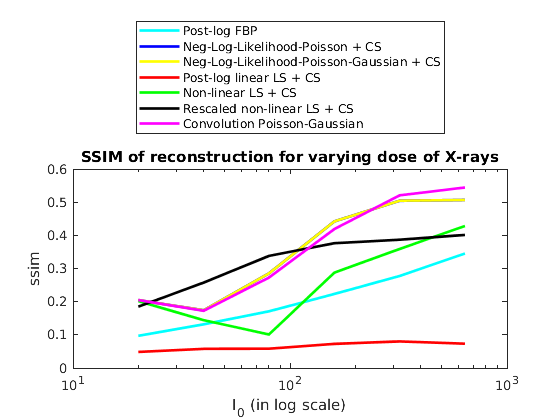}
       \end{subfigure}
      \begin{subfigure}[b]{0.45\linewidth}
        \includegraphics[width=\textwidth]{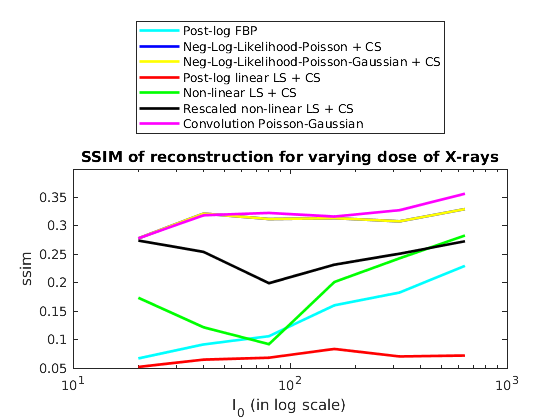}
      \end{subfigure}
\caption[SSIM of reconstructions for varying values of X-ray doses of Colon.]{SSIM of the reconstructions for Walnut and Colon datasets shown in Fig.~\ref{fig:colon_result} for varying values of X-ray doses. A higher SSIM implies better reconstruction. Here, the reconstructions by Poisson-likelihood and Poisson-Gaussian likelihood methods were very similar. Hence, their SSIM plots (blue and yellow respectively) overlap.}
\label{fig:low_dose_comparison_SSIM}
\end{figure}

Sample reconstructions are shown in  Figs.~\ref{fig:walnut_result} and~\ref{fig:colon_result}. The corresponding SSIM values of the reconstructions are shown in Fig.~\ref{fig:low_dose_comparison_SSIM}. From these plots, the following can be inferred: the convolution method and the Poisson-Gaussian likelihood reconstructions were comparable and gave the best reconstructions for a majority of dose levels and datasets.The Poisson-Gaussian Likelihood and the Poisson-only likelihood have very similar performance. However, at a theoretical level, the former is a more principled method, and can deal with negative-valued measurements which have to be weeded out for the Poisson-only method. The non-linear least squares method (Sec.~\ref{subsubsec:NLLSCS}) performed poorly. This is because the data-fidelity term assumes constant variance for all signal values. In reality, the variance of Poisson noise increases as signal intensity increases. The post-log linear least squares (Sec.~\ref{subsubsec:post_log_CS}) failed because the linear model fails to approximate the highly non-linear low-dose acquisition. The post-log FBP yielded poor results, especially at slightly higher dose levels (for example at $I_0 = 620$ in Fig.~\ref{fig:walnut_result}. This could be due to the absence of iterative optimization when compared to the other methods and due to the post-log approximation. For all datasets except Walnut (Colon as discussed here, and Pelvis, Shoulder as discussed in~\cite{videos}), the performance of \textit{rescaled non-linear least squares (RNLLS)} is inbetween the performance of likelihood-based methods and those of all other methods. For the Walnut dataset though, the RNLLS gives the best quality for many dosage levels. The performance of the above methods across multiple noise instances is discussed in Sec.2.1 of~\cite{videos}. 

\section{Reconstruction with prior}
\label{sec:low_dose_reconstruction_with_prior}

As seen so far, principled data fidelity terms play a significant role in improving the reconstruction performance. However, when the x-ray dose is less, the performance can be further improved by incorporation of useful priors~\cite{PICCS,Rashed2016}. These priors could be previous high-quality reconstructions of the same object in longitudinal studies, or high-quality reconstructions of similar objects. We refer to such prior data as templates.  Here, our aim is to reconstruct an object from its \emph{low-dose measurements}, using templates which are previous \emph{high-dose reconstructions} of the same object in a longitudinal study. However, there is a danger of the templates overwhelming the current reconstruction and adversely affecting reconstruction of new regions in the test (i.e., the object which needs to be reconstructed from the current set of new tomographic projections) that are absent in any of the templates. In the case of reconstruction from few projection views, the above problem was tackled~\cite{Preeti2018} by generating a map (known as `weights-map') that shows an estimate of the regions of new changes and their magnitude. This map was then used to modulate the influence of the prior on the reconstruction of the test. The weights-map was computed based on the difference between the pilot reconstruction from the test measurements and its projection onto an eigenspace spanned by representative templates. However, in the low-dose case, this is not a preferable method because \textit{all information about the noise model is valid for the measurement space alone. The noise model (i.e., $\boldsymbol{y} \sim \rm{Poisson}(I_0\exp\{-\boldsymbol{\Phi x}\}) + \boldsymbol{\eta}$) is not applicable to the spatial reconstructed image domain.} 

Hence, in this work, we propose a new algorithm to compute the weights-map (i.e to detect differences between the test and the templates) directly in the measurement space. The aim is to identify those measurement bins which correspond to the new changes in the test. Following are the steps followed in order to accomplish this:

\begin{enumerate}
\item Let $\boldsymbol{x_{t_1}}$...$\boldsymbol{x_{t_n}}$ be $n$ high quality template volumes, i.e. template volumes reconstructed from their standard dose measurements.
\item Simulate noiseless measurements from template volumes using the same $I_0$ used for imaging the test i.e.  $\boldsymbol{y_{t_i}} = I_0\exp\{-\boldsymbol{\Phi x_{t_i}}\} $, where $1\leq i \leq n$.
\item Let $\boldsymbol{y_{t_{i,j}}}$ be the tomographic projection of the $i^{th}$ template from the $j^{th}$ angle, where $1 \leq j \leq Q$. Let $\{E_j\}_{j=1}^Q$ represent the set of eigenspaces, where $E_j$ is the eigenspace built from the tomographic projections of each of the templates in the $j^{th}$ angle, i.e. built from $\{\boldsymbol{y}_{t_{i,j}}\}_{i=1}^n$
\item Let $\boldsymbol{y}_j$ be the noisy tomographic projection of the test volume $\boldsymbol{x}$ from the $j^{th}$ angle. For each $j \in \{1,...,Q\}$, project $\boldsymbol{y}_j$ onto $E_j$, i.e.,
 compute the eigen-coefficients  $\boldsymbol{\alpha}^m_j$ of the measurements  $\boldsymbol{y}_j$, along the set of eigenvectors $\boldsymbol{V}^m_j$:
    \begin{equation}
      \boldsymbol{\alpha}^m_j = (\boldsymbol{V}^m_j)^T(\boldsymbol{y}_j -\boldsymbol{\mu}^m_j)
      \end{equation}
        where $\boldsymbol{\mu}^m_j$ denotes the mean tomographic projection of all templates in the $j^{th}$ angle. The $m$ in the suffix denotes that the eigenspace  $E_j := \{\boldsymbol{\mu}^m_j,\boldsymbol{V}^m_j\}$  is computed in the measurement space (We will contrast this with another eigen-space computed in image domain,  used later in Eq.~\ref{eq:low_dose_weighted_prior}). Next, compute the resultant projection $\boldsymbol{y}_{pj}$, i.e.,
        \begin{equation}
          \boldsymbol{y}_{pj} = \boldsymbol{\mu}^m_j + \boldsymbol{V}^m_j\boldsymbol{\alpha}^m_j
         \end{equation}
\item Note that if a random variable $s \sim Poisson(\lambda) + \eta$, where $\eta \sim \mathcal{N}(0,\sigma^2)$, then 
$\sqrt{s + (3/8) + \sigma^2}$ is approximately distributed as $N(\sqrt{\lambda + (3/8) + \sigma^2},1/4)$. The quality of the approximation is known to improve as $\lambda$ increases. In the absence of Gaussian noise (equivalent to the case where $\sigma = 0$), this transform is called the Anscombe transform~\citep{anscombe1948,Curtiss1943}, and has been widely used in image processing. In the presence of Gaussian noise, it is referred to as the generalized Anscombe transform~\citep{Murtagh1995}. Now consider the $k^{th}$ bin in the test measurement $\boldsymbol{y}$ as well as in $\boldsymbol{y_{pj}}$, which we shall denote as $y(k)$ and $y_{p_j}(k)$ respectively. If $y(k)$ represents the same underlying structure as in $y_{p_j}(k)$, barring the effect of Poisson-Gaussian noise, i.e. if the $k^{th}$ bin in $\boldsymbol{y}$ is not part of the `new changes', then the following is true:

$\sqrt{\boldsymbol{y} + 3/8 + \sigma^2} - \sqrt{\boldsymbol{y_p} + 3/8 + \sigma^2} \sim N(0,1/4)$.

For bins falling in the regions of change in the test (compared to the template projections), the above hypothesis is false. The same argument can be extended for entire segments or 2D regions.

\item Based on the aforementioned fact, hypothesis testing is performed on $\sqrt{\boldsymbol{y} + 3/8 + \sigma^2} - \sqrt{\boldsymbol{y_p} + 3/8 + \sigma^2}$ to detect bins corresponding to new changes in the measurement space. We use Z-test for hypothesis testing on 2D patches in the measurement space (note that since the volume is in 3D, the measurement space is in 2D for every imaging view). This $Z$ test computes the probability that the given sample is likely to be drawn from a population as specified by the null hypothesis. In this case, the null hypothesis is that the intensity values of the patches are drawn from $\mathcal{N}(0,1/4)$. The confidence level was set to $95\%$, i.e. for null hypothesis to be false, the probability $p$ that the sample is drawn from Normal distribution must lie in the $2.5\%$ tail-end of the Normal distribution on either side. A lower $p$-value denotes the presence of new changes i.e., presence of differences between the test and the templates in the measurement bins.
\item Once the new changes are detected in the measurement space, filtered backprojection of the vectors (containing p-values) resulting from the hypothesis test gives the location of the new changes (which we denote $W_{inlier}$) in the original (3D) spatial domain. The Cosine filter was used in the filtered backprojection process.
\item The final weights-map $\boldsymbol{W}$ \footnote{An alternate method to compute a weights-map (a simpler binary weights-map) is discussed in Sec.3 of~\cite{videos}} is computed from $\boldsymbol{W}_{inlier}$ by the following steps:
(a) \textbf{Inversion}: $\boldsymbol{W} = 1./(1 + (\boldsymbol{W}_{inlier}).^2)$. This step is just for inversion so that new regions get lower weight/intensity than prior-similar regions, (b) \textbf{Linear stretching}: Perform linear stretching on $\boldsymbol{W}$ so that the weights lie between 0 and 1.
\end{enumerate}

 Finally, the computed weights-map is used in a reconstruction optimization as follows:
\begin{equation}
J(\boldsymbol{\theta},\boldsymbol{\alpha}) = \sum_{i=1}^m\frac{(\boldsymbol{y}_i - I_0e^{(-\boldsymbol{\Phi\Psi\theta})_i})^2}{I_0e^{(-\boldsymbol{\Phi\Psi\theta})_i} + \sigma^2} + \lambda_1\lVert\boldsymbol{\theta}\rVert_1 +\lambda_2\lVert\boldsymbol{W}(\boldsymbol{x} - (\boldsymbol{\mu} + \sum_{i}\boldsymbol{V_i}\alpha_i))\rVert_2^2
\label{eq:low_dose_weighted_prior}
\end{equation}
where the eigenvectors $\boldsymbol{V}$ and mean of the templates $\boldsymbol{\mu}$ form the eigenspace which  is built from the available high-dose \emph{reconstructions} of the templates. $\boldsymbol{\alpha}$ denotes the coefficients of $\boldsymbol{V}$, when the pilot reconstruction of the test is projected onto this eigenspace. Information about the location and magnitude of new changes in the test is present in the weights-map $\boldsymbol{W}$.  Eq.~\ref{eq:low_dose_weighted_prior} is solved by alternating minimization on $\boldsymbol{\theta}$ and $\boldsymbol{\alpha}$ until convergence is reached. 
\subsection{Reconstruction results}

\begin{figure}[!b]
\centering
    \begin{subfigure}[b]{0.19\linewidth}
        \includegraphics[width=\textwidth]{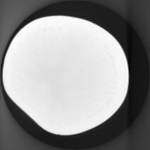}
\captionsetup{labelformat=empty}       
 \caption{}
    \end{subfigure}
    \begin{subfigure}[b]{0.19\linewidth}
        \includegraphics[width=\textwidth]{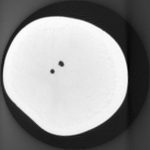}
\captionsetup{labelformat=empty}
        \caption{}
     \end{subfigure}
    \begin{subfigure}[b]{0.19\linewidth}
        \includegraphics[width=\textwidth]{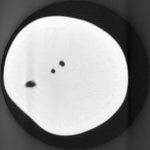}
\captionsetup{labelformat=empty}
        \caption{}
     \end{subfigure}
    \begin{subfigure}[b]{0.19\linewidth}
        \includegraphics[width=\textwidth]{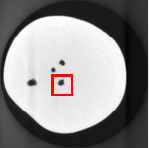}
\captionsetup{labelformat=empty}
        \caption{}
     \end{subfigure}
     \caption[Templates chosen for prior-based low-dose reconstruction on Potato]{Potato 3D dataset: One of the slices from template volumes (first four from the left) and test volume (extreme right). Size of each volume is [$150 \times 150 \times 20$].}
\label{fig:low_dose_potato3D_templates_test}
\end{figure}
\begin{figure}[!t]
\centering
\subcaptionbox{Test}{\includegraphics[width=0.19\columnwidth]{images/low_dose_prior/3D/potato/testIm_roi.png}}\hfill
\subcaptionbox{No prior}{\includegraphics[width=0.19\columnwidth]{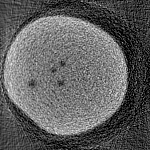}}\hfill
\subcaptionbox{Unweighted}{\includegraphics[width=0.19\columnwidth]{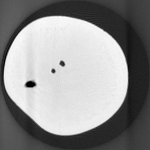}}\hfill
\subcaptionbox{Our\\reconstruction}{\includegraphics[width=0.19\columnwidth]{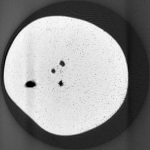}}
\subcaptionbox{Weights $W$\label{fig:potato_weights}}{\includegraphics[width=0.19\columnwidth]{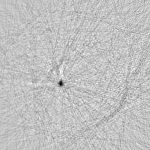}}
\caption[Prior-based low-dose reconstruction on Potato 3D dataset]{Prior-based low-dose reconstruction on 3D potato dataset. (a) Slice from test volume (b) Reconstruction using no prior (using RNLLS of Sec.~\ref{sec:rescaledNLLSCS}); $SSIM = 0.22$ (c) Slice from unweighted prior reconstruction; $SSIM = 0.42$ . The new change is missing. (d) Slice from weighted prior reconstruction; $SSIM = 0.69$. The new change is detected here and its reconstruction is guided by the low-dose measurements. (e) Weights map showing the location and intensity of the new changes. All SSIM values are averaged over 14 slices of the reconstructed volume in the red RoI region. The reconstructed volumes can be seen in~\citep{videos}.}
\label{fig:low_dose_potato3D_results}
\end{figure}

The above algorithm was validated by reconstructing a 3D volume from its low dose measurements. Fig.~\ref{fig:low_dose_potato3D_templates_test} shows a slice from each of the template and test volumes of the potato dataset. This dataset~\footnote{We are grateful to Dr.~Andrew Kingston for facilitating data collection at the Australian National University.} consisted of four scans of the humble potato, chosen for its simplicity. Measurements from each scan consisted of cone-beam projections from 900 views, each of size $150 \times 150$. The corresponding size of the
reconstructed volume is $150\times150\times150$. While the first scan was taken of the undistorted
potato, subsequent scans were taken of the same specimen, each time after drilling a new
hole halfway into the potato. The ground truth consists of FDK reconstructions
from the full set of acquired measurements from $900$ projection views. Low dose cone-beam measurements were simulated from full-view FDK reconstructions of the test volume. $I_0$ was set to 4000, a value corresponding to Poisson noise of $1.5\%$. Mean of the added Gaussian noise was $0$ and $\sigma$ was set to $0.1\%$ of the mean of Poisson-corrupted measurements. Fig~\ref{fig:low_dose_potato3D_results} shows the same slice from each of the reconstructed volumes. A patch size of $[5,5]$ was used for hypothesis testing and the location of new changes (marked in red RoI in test) was accurately detected in the weights-map as seen in Fig.~\ref{fig:potato_weights}. The reconstructed volumes can be found in~\citep{videos}.

\subsection{Re-irradiation to improve reconstruction}
\begin{figure}[!h]
    \begin{subfigure}[b]{0.19\linewidth}
        \includegraphics[width=\textwidth]{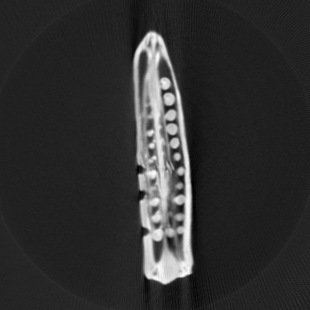}
\captionsetup{labelformat=empty}       
 \caption{}
    \end{subfigure}
    \begin{subfigure}[b]{0.19\linewidth}
        \includegraphics[width=\textwidth]{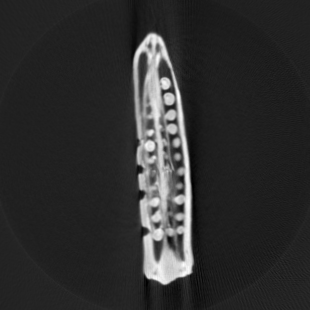}
\captionsetup{labelformat=empty}
        \caption{}
     \end{subfigure}
    \begin{subfigure}[b]{0.19\linewidth}
        \includegraphics[width=\textwidth]{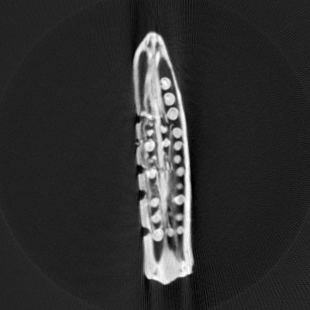}
\captionsetup{labelformat=empty}
        \caption{}
     \end{subfigure}
    \begin{subfigure}[b]{0.19\linewidth}
        \includegraphics[width=\textwidth]{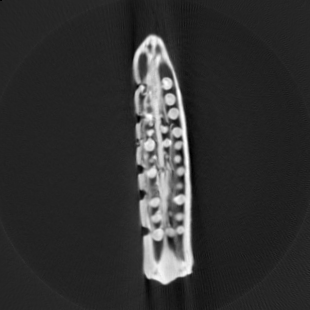}
\captionsetup{labelformat=empty}
        \caption{}
     \end{subfigure}
    \begin{subfigure}[b]{0.19\linewidth}
        \includegraphics[width=\textwidth]{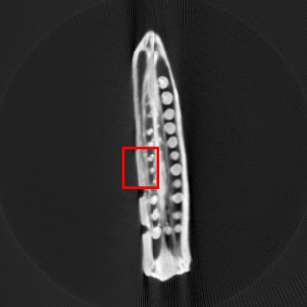}
\captionsetup{labelformat=empty}
        \caption{}
     \end{subfigure}
     \caption[Dataset for re-irradiation]{Dataset for illustrating re-irradiation: Templates (first four from the left) and test (extreme right). Size of each slice is $(310\times310)$. The RoI shows the region of difference between the test and the templates.}
\label{fig:re-irradiation_templates_test_okra}
\end{figure}

\begin{figure}[!h]
\centering
\subcaptionbox{Test}{\includegraphics[width=0.19\columnwidth]{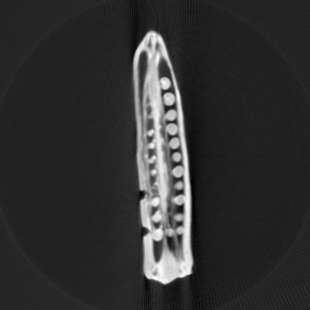}}\hfill
\subcaptionbox{Pilot}{\includegraphics[width=0.19\columnwidth]{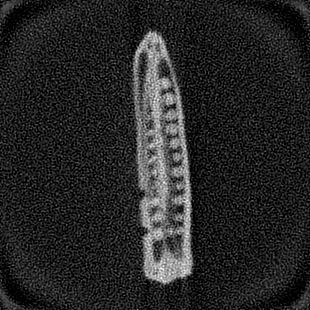}}\hfill
\subcaptionbox{weights $W$}{\includegraphics[width=0.19\columnwidth]{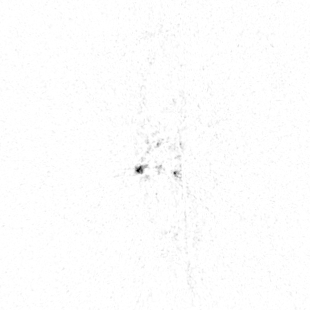}}\hfill
\subcaptionbox{Weighted\\ Prior}{\includegraphics[width=0.19\columnwidth]{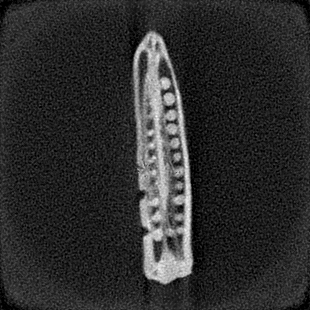}}\hfill
\subcaptionbox{After\\ re-irradiation}{\includegraphics[width=0.19\columnwidth]{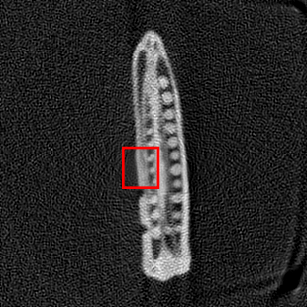}}
\caption[Re-irradiation results in Okra]{Improving reconstruction by re-irradiation in Okra 2D dataset. (a) test (b) pilot (c) weights-map; the lower the intensity, the higher the magnitude of new changes. (d) weighted prior reconstruction; the quality of reconstruction of new regions is poor because it is guided by the measurements alone. (e) re-irradiated reconstruction; \emph{new measurements with twice the earlier low-dose X-ray intensity at $20\%$ of the bins enable better reconstruction of new regions} (as shown in RoI).}
\label{fig:re-irradiation_okra_results}
\end{figure}

\begin{figure}[!h]
    \begin{subfigure}[b]{0.16\linewidth}
        \includegraphics[width=\textwidth]{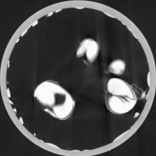}
\captionsetup{labelformat=empty}       
 \caption{}
    \end{subfigure}
    \begin{subfigure}[b]{0.16\linewidth}
        \includegraphics[width=\textwidth]{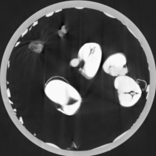}
\captionsetup{labelformat=empty}
        \caption{}
     \end{subfigure}
    \begin{subfigure}[b]{0.16\linewidth}
        \includegraphics[width=\textwidth]{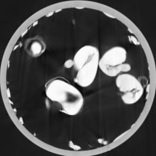}
\captionsetup{labelformat=empty}
        \caption{}
     \end{subfigure}
    \begin{subfigure}[b]{0.16\linewidth}
        \includegraphics[width=\textwidth]{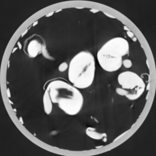}
\captionsetup{labelformat=empty}
        \caption{}
    \end{subfigure}
        \begin{subfigure}[b]{0.16\linewidth}
        \includegraphics[width=\textwidth]{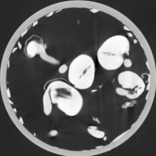}
\captionsetup{labelformat=empty}
        \caption{}
     \end{subfigure}
    \begin{subfigure}[b]{0.16\linewidth}
        \includegraphics[width=\textwidth]{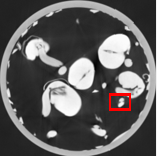}
\captionsetup{labelformat=empty}
        \caption{}
     \end{subfigure}
     \caption[Sprouts Dataset for re-irradiation]{Sprouts Dataset for illustrating re-irradiation: Templates (first row) and test (second row). Size of each slice is $(156\times156)$. The RoI shows the region of difference between the test and the templates.}
\label{fig:re-irradiation_templates_test_sprouts}
\end{figure}

\begin{figure}[!h]
\centering
\subcaptionbox{Test}{\includegraphics[width=0.19\columnwidth]{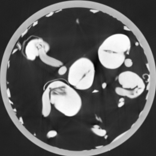}}\hfill
\subcaptionbox{Pilot}{\includegraphics[width=0.19\columnwidth]{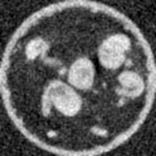}}\hfill
\subcaptionbox{weights $W$}{\includegraphics[width=0.19\columnwidth]{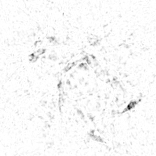}}\hfill
\subcaptionbox{Weighted\\ Prior}{\includegraphics[width=0.19\columnwidth]{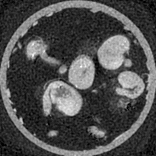}}\hfill
\subcaptionbox{After\\ re-irradiation}{\includegraphics[width=0.19\columnwidth]{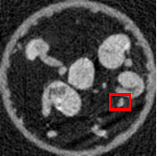}}
\caption[Re-irradiation results in Sprouts]{Improving reconstruction by re-irradiation in Sprouts 2D dataset. (a) test (b) pilot (c) weights-map; the lower the intensity, the higher the magnitude of new changes. (d) weighted prior reconstruction; the quality of reconstruction of new regions is poor because it is guided by the measurements alone. (e) re-irradiated reconstruction; \emph{new measurements with $8$ times the earlier low-dose X-ray intensity at $25\%$ of the bins enable better reconstruction of new regions} (as shown in RoI).}
\label{fig:re-irradiation_sprouts_results}
\end{figure}

Once the regions of new changes are detected by the weights map, this information can be used to re-irradiate them  with standard-dose rays and further improve the quality of their reconstruction. Following are the steps of the re-irradiation process: 

\begin{enumerate}
\item Let the X-rays passing through the new regions have their source points denoted by $S_1$, and the corresponding bins at the detector be denoted by $D_1$. Let the X-rays passing through the other regions (i.e. regions where the test and the templates are not structurally different) have their source points denoted by $S_2$, and the corresponding bins at the detector be denoted by $D_2$.
\item Block $S_2$ and re-irradiate the object by passing standard-dose rays from $S_1$. This will generate measurements of high quality for regions of new changes. If the regions of new change are small in area, this process incurs only a small cost for the extra amount of radiation, since the latter is restricted to only specific regions. 
\item In the measurement matrix captured for pilot reconstruction, replace all the bins in set $D_1$ by their new measurements. Therefore, the final measurement matrix consists of standard-dose measurements corresponding to new regions of the object and low-dose measurements corresponding to the other regions of the object.
\end{enumerate}
The new measurement model is: 
$\boldsymbol{y} \sim \rm{Poisson}(\boldsymbol{I_0}\exp\{-\boldsymbol{\Phi x}\}) + \boldsymbol{\eta}$.
Here $\boldsymbol{I_0}$ denotes a diagonal matrix with $\boldsymbol{I_0}(k,k)$ denoting the strength of the X-ray incident on the $k^{th}$ bin of the sensor. Fig.~\ref{fig:re-irradiation_templates_test_okra} shows the templates and test images, and Fig.~\ref{fig:re-irradiation_okra_results} shows the reconstructions illustrating the benefit of re-irradiation. The new changes within the RoI are reconstructed very well after they are re-imaged with standard-dose X-rays. This is also reinforced by results on the sprouts data (Fig.~\ref{fig:re-irradiation_templates_test_sprouts}), shown in Fig.~\ref{fig:re-irradiation_sprouts_results}. The selection of bins for re-irradiation and the choice of new X-ray intensity can also be chosen in a supervised manner by the physician or scientist based on the particular clinical or non-clinical setting.

\section{Tuning of parameters}
\label{sec:tuning_hyper_parameters}
Two parameters were used in the techniques presented in this chapter: $\lambda_1$: weight for CS term and $\lambda_2$: weight for object-prior. Below are few of the ways to select these parameters.
\subsection{Selection of weightage for CS term}

In a large body of work on tomographic reconstruction~\citep{Zheng2016},~\citep{liu2016}, the regularization parameter $\lambda_1$ is chosen in an ``omniscient fashion". That is, the optimization problem is solved separately for many different values of $\lambda_1$. The particular result which yields the least MSE with respect to a ground truth image is chosen to be the correct result. Such a method requires knowledge of the ground truth, and hence is infeasible in practice. Other alternatives include visual inspection or cross-validation. However none of these techniques are fully practical. Instead, we propose a method to choose $\lambda$ based on sound statistical principles pertaining to the Poisson or the Poisson-Gaussian noise model. The method is shown here in conjunction with the rescaled non-linear least squares method, however in principle, it can be used with any data fidelity term. For the Poisson-Gaussian noise model, the cost function is given by
$J(\boldsymbol{\theta}) = \sum_{i=1}^m\frac{(y_i - I_0e^{(-\boldsymbol{\Phi\Psi\theta})_i})^2}{I_0e^{(-\boldsymbol{\Phi\Psi\theta})_i}+ \sigma^2} + \lambda_1\lVert\boldsymbol{\theta}\rVert_1$.

\begin{figure}[!h]
    \begin{subfigure}[b]{0.48\linewidth}
        \includegraphics[width=\textwidth]{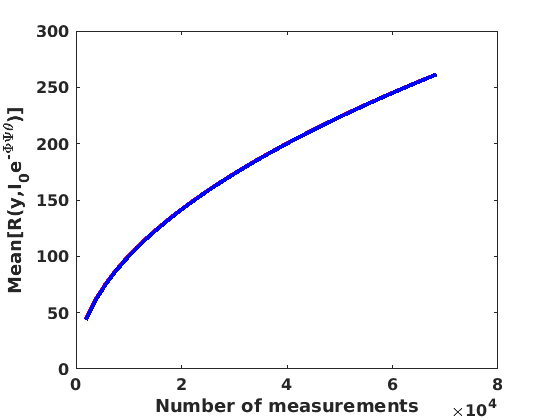}
 \caption{}
    \end{subfigure}
    \begin{subfigure}[b]{0.48\linewidth}
        \includegraphics[width=\textwidth]{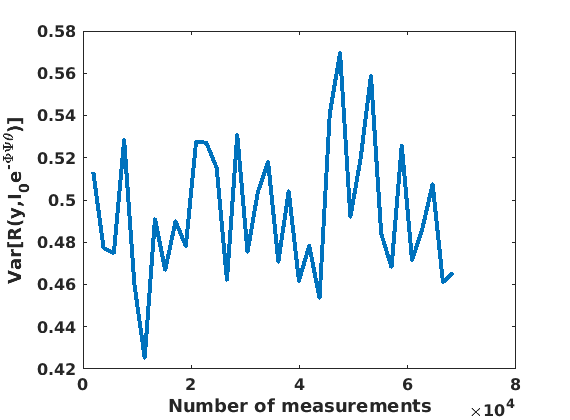}
        \caption{}
     \end{subfigure}
    \begin{subfigure}[b]{0.48\linewidth}
        \includegraphics[width=\textwidth]{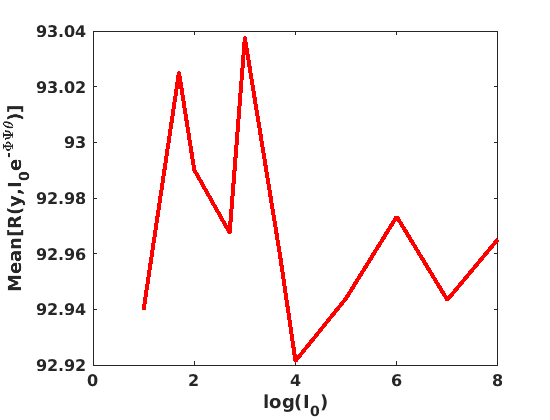}
        \caption{}
     \end{subfigure}
    \begin{subfigure}[b]{0.48\linewidth}
        \includegraphics[width=\textwidth]{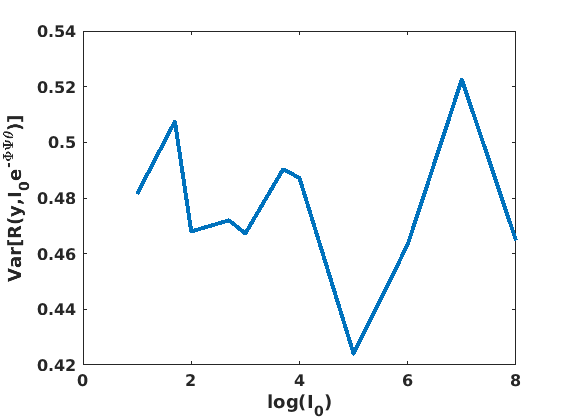}
        \caption{}
    \end{subfigure}
     \caption[Statistics of rescaled data fidelity]{Mean and variance of the data-fidelity term $R =  \sum_{i=1}^m\frac{(y_i - I_0e^{(-\boldsymbol{\Phi\Psi\theta})_i})^2}{I_0e^{(-\boldsymbol{\Phi\Psi\theta})_i}+ \sigma^2}$ for different number of measurements (projection views) and beam strength $I_0$. (a) Expected value of $R$ exactly coincides with $\sqrt{m}$, (b) Variance of $R$ is insignificant for any number of measurements, (c) mean of $R$ is independent of the beam-strength, and (d) Variance of $R$ is insignificant for all $I_0$ values.}
\label{fig:statistics}
\end{figure}

\begin{figure}[h!]
    \begin{subfigure}[b]{0.19\linewidth}
        \includegraphics[width=\textwidth]{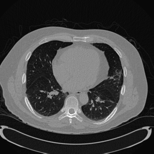}
\captionsetup{labelformat=empty}       
 \caption{Test}
    \end{subfigure}
    \begin{subfigure}[b]{0.19\linewidth}
        \includegraphics[width=\textwidth]{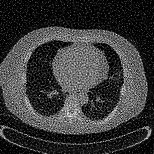}
\captionsetup{labelformat=empty}       
 \caption{$\lambda = 0.0001$}
    \end{subfigure}
     \begin{subfigure}[b]{0.19\linewidth}
        \includegraphics[width=\textwidth]{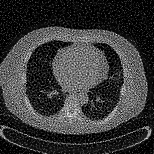}
\captionsetup{labelformat=empty}       
 \caption{$\lambda = 0.0010$}
    \end{subfigure}
       \begin{subfigure}[b]{0.19\linewidth}
        \includegraphics[width=\textwidth]{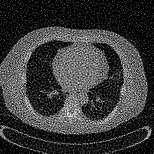}
\captionsetup{labelformat=empty}       
 \caption{$\lambda = 0.01$}
    \end{subfigure}
       \begin{subfigure}[b]{0.19\linewidth}
        \includegraphics[width=\textwidth]{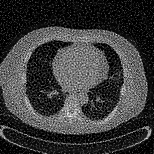}
\captionsetup{labelformat=empty}       
 \caption{$\lambda = 0.10$}
    \end{subfigure}
       \begin{subfigure}[b]{0.19\linewidth}
        \includegraphics[width=\textwidth]{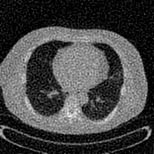}
\captionsetup{labelformat=empty}       
 \caption{$\lambda = 1.00$}
    \end{subfigure}
       \begin{subfigure}[b]{0.19\linewidth}
        \includegraphics[width=\textwidth]{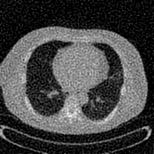}
\captionsetup{labelformat=empty}       
 \caption{$\lambda = 1.10$}
    \end{subfigure}
       \begin{subfigure}[b]{0.17\linewidth}
         \fcolorbox{green}{green}{\includegraphics[width=\textwidth]{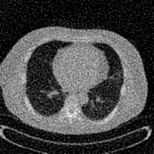}}
\captionsetup{labelformat=empty}       
 \caption{$\lambda = 1.20$}
    \end{subfigure}
      \begin{subfigure}[b]{0.19\linewidth}
        \includegraphics[width=\textwidth]{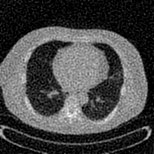}
\captionsetup{labelformat=empty}       
 \caption{$\lambda = 1.30$}
    \end{subfigure}
      \begin{subfigure}[b]{0.19\linewidth}
        \includegraphics[width=\textwidth]{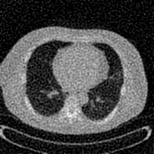}
\captionsetup{labelformat=empty}       
 \caption{$\lambda = 1.400$}
    \end{subfigure}
      \begin{subfigure}[b]{0.17\linewidth}
         \fcolorbox{red}{red}{\includegraphics[width=\textwidth]{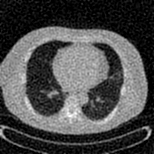}}
\captionsetup{labelformat=empty}       
\caption{$\lambda = 2.00$}
      \end{subfigure}
      \quad
      \begin{subfigure}[b]{0.19\linewidth}
        \includegraphics[width=\textwidth]{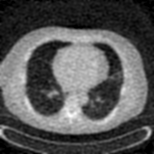}
\captionsetup{labelformat=empty}       
 \caption{$\lambda = 5.00$}
    \end{subfigure}
      \begin{subfigure}[b]{0.19\linewidth}
        \includegraphics[width=\textwidth]{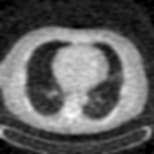}
\captionsetup{labelformat=empty}       
 \caption{$\lambda = 10.0$}
    \end{subfigure}
      \begin{subfigure}[b]{0.19\linewidth}
        \includegraphics[width=\textwidth]{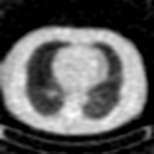}
\captionsetup{labelformat=empty}       
 \caption{$\lambda = 15.0$}
    \end{subfigure}
      \begin{subfigure}[b]{0.19\linewidth}
        \includegraphics[width=\textwidth]{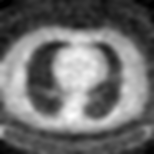}
\captionsetup{labelformat=empty}       
 \caption{$\lambda = 20.0$}
    \end{subfigure}
     \caption[Colon Reconstructions for varying $\lambda_1$ ]{Colon test data and its reconstructions for different values of $\lambda_1$. $D$ is minimum for $\lambda_1 = 1.2$, shown in green, with a relative MSE of $0.1691$. The reconstruction for $\lambda_1 = 2$, shown in red,  gives the minimum relative MSE of $0.1501$.}
\label{fig:colon_discrepancy_results}
\end{figure}

Let $m$ denote the total number of bins, $\boldsymbol{\theta_{opt}}$ the reconstruction with optimal $\lambda_1 = \lambda_{1\_opt}$. Let $a_i \triangleq I_o e^{-[\boldsymbol{\Phi \Psi \theta_{opt}]}_i}$. Clearly, we have $Var(y_i) = a_i + \sigma^2$. Hence we can state that
 $E[\sum_{i=1}^m (y_i - a_i)^2/ (a_i + \sigma^2)] = m$.
Furthermore, our simulations (Fig.~\ref{fig:statistics}) have shown that
\begin{equation}
  E\Big(\lVert(y-I_0 e^{-\boldsymbol{\Phi \Psi \boldsymbol{\theta_{opt}}}}) \oslash \sqrt{I_0 e^{-\boldsymbol{\Phi \Psi \boldsymbol{\theta_{opt}}}} + \sigma^2}\rVert_2\Big) \approx \sqrt{m}
\end{equation}
where $\oslash$ denotes element-wise division. We also observed that the variance of the above quantity is very small. This is illustrated in Fig.~\ref{fig:statistics}, which shows that the variance of $R =  \sum_{i=1}^m\frac{(y_i - I_0e^{(-\boldsymbol{\Phi\Psi\theta})_i})^2}{I_0e^{(-\boldsymbol{\Phi\Psi\theta})_i}+ \sigma^2}$ is very small compared to its mean. The expected value of  $R$ varies with the number of measurements (is equal to $\sqrt{m})$, and is independent of $I_0$. Hence we conclude that the quantity $|| (y-I_0 e^{-\boldsymbol{\Phi \Psi \theta_{opt}}}) \oslash\sqrt{I_0 e^{-\boldsymbol{\Phi \Psi \theta_{opt}}}+ \sigma^2} ||_2$ should be as close to $\sqrt{m}$ as possible. Therefore, we consider  
\begin{equation}
D = \textrm{abs}\Big(\big\lVert (\boldsymbol{y} - I_0e^{\boldsymbol{-\Phi\Psi\theta_{opt}}}) \oslash \sqrt{(I_0e^{\boldsymbol{-\Phi\Psi\theta_{opt}}} + \sigma^2)}\big\rVert_2- \sqrt m\Big)
\end{equation}
\begin{figure}[h!]
\centering
  \includegraphics[width=0.6\textwidth]{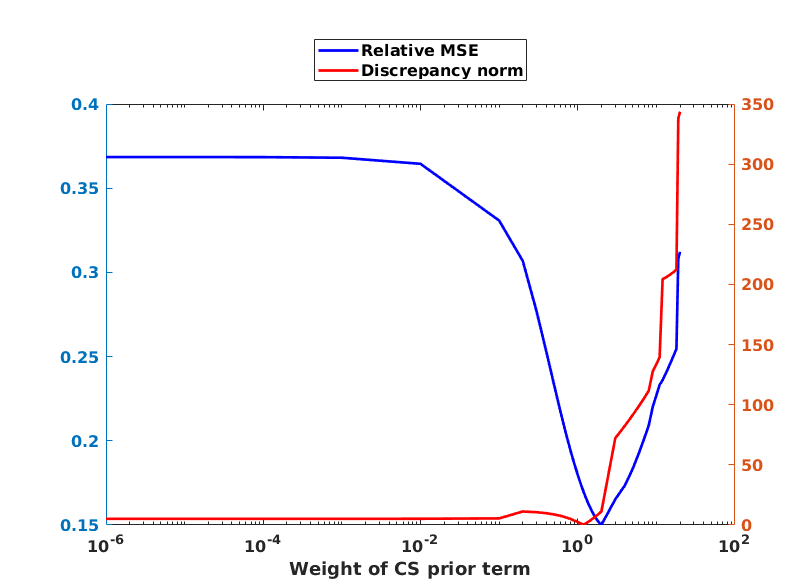}
  \caption[A method to select $\lambda_1$]{A method to choose the parameter $\lambda_1$ in low-dose reconstruction: We expect $D$ to be minimum at the same $\lambda_1$ for which relative MSE is minimum. Here, the $\lambda_1$ for which $D$ and relative MSE are minimum are very close. Refer to Fig.~\ref{fig:colon_discrepancy_results} to observe the reconstruction results for different values of $\lambda_1$.}
  \label{fig:discrepancy_rescaled_plot}
\end{figure}
and observe how $D$ and relative MSE of reconstructions vary for different values of $\lambda_1$. At the optimum $\lambda_1$, $D$ must be minimum. The test image ($154\times154$) and the reconstructions are shown in Figure~\ref{fig:colon_discrepancy_results}. For these reconstructions, 410 projection views were chosen and Gaussian noise = $0.3\%$ was added to the measurements. The dose of X-rays resulted in a Poisson NSR of 0.018. As shown in Fig.~\ref{fig:discrepancy_rescaled_plot}, the $\lambda_1$ for which $D$ and relative MSE are minimum, are very close. In a real-life setting, when relative MSE cannot be computed because of absence of ground-truth, a brute force search needs to be done followed by selecting the value of $\lambda_1$ that minimizes $D$.

\subsection{Selection of weightage for object-prior term}
The weightage for the object prior, $\lambda_2$ term needs to be chosen omnisciently for every dataset. However, for a variation of $\pm 300$, there was no significant effect on the reconstructions. Lower values indicate that the reconstructions are primarily guided by the measurements, and higher values will strengthen the effect of the prior.

\section{Conclusions}
\label{sec:conclusions}
 In the low-dose CT imaging regime, the noise in the measurements becomes significant and needs to be accounted for during the reconstruction. Two new techniques: Poisson-Gaussian convolution and rescaled non-linear least squares (RNLLS) were presented and extensively compared with many of the existing methods. RNLLS was further used in low-dose reconstruction for longitudinal studies to specifically detect new regions in the test and simultaneously reduce noise in the other reconstructed regions. The results were validated on both 2D and 3D biological data. We demonstrated that the reconstructions of the regions of new changes can be significantly improved by re-irradiating these specific regions by standard-dose X-rays. Further, different methods for choosing the parameters $\lambda_1, \lambda_2$ were also discussed, which has not been dealt with in literature. Our technique can possibly be extended to the case where templates of a similar class of objects are available, as against previous scans of the same object. This may further increase the utility of the technique in clinical settings.

\small{\bibliography{lowdose}}
\end{document}